%Paper: hep-th/9308031
%From: Roberto Percacci <PERCACCI@sbitp.itp.ucsb.edu>
%Date: Sat, 7 Aug 1993 03:08 PST

\def\eff{{\rm eff}}
\def\dslash{D}

\def\o{{\cal O}}
\def\bo{{\bar{\cal O}}}
\def\to{{\tilde{\cal O}}}

\font\titlefont=cmbx10 scaled\magstep1
\magnification=\magstep1
\null
\rightline{ITP XXXXXX}
\rightline{SISSA xx/93/EP}
\rightline{hep-th/9308031}
\vskip 1.5cm
\centerline{\titlefont THE EFFECTIVE POTENTIAL}
\centerline{\titlefont FOR THE CONFORMAL FACTOR}
\centerline{\titlefont IN THE STANDARD MODEL AND BEYOND}
\smallskip
\vskip 1.5cm
\centerline{\bf R. Percacci \footnote{$^{*}$}
{\it permanent address: \rm SISSA, via Beirut 4, 34014 Trieste, Italy}}
\smallskip
\centerline{Institute for Theoretical Physics}
\centerline{University of California, Santa Barbara, CA 93106-4030}
\vskip 2.2cm
\centerline{\bf Abstract}
\smallskip\midinsert\narrower\narrower\noindent
There is a general mechanism by which certain matter fields coupled to
gravity can generate a nontrivial effective potential for the conformal
factor of the metric.
It is based on a nonstandard regularization method, with the cutoff
being defined independently of the conformal factor.
This mechanism produces a coupling of the matter fields to a dilaton,
and a complicated interaction between matter, dilaton and metric.
When it is applied to the standard model, it gives an effective potential
which can be used to predict the top and Higgs masses.
If the purely gravitational contribution to the potential is
added, the mass of the dilaton is of the order of Planck's mass
and the large hierarchy between the Planck and Fermi scales
appears to be due to the smallness of the Higgs-dilaton coupling.
\endinsert
\vskip 1cm
\vfil\eject
It has been conjectured that the existence of a scalar dilaton
might be related to the vanishing of the cosmological constant [1].
In particular, this has prompted studies of the coupling of a
dilaton to the standard model [2]. Very recently, a new twist has
been added to the subject: the existence of a dilaton and its
classical coupling to the standard model have been derived from
arguments of noncommutative geometry [3] and then, modulo some
assumptions about the structure of radiative corrections,
a very sharp prediction of the top quark mass was derived [4].
It is quite clear that the only role played by noncommutative
geometry in this argument is to make a certain statement about
the effective coupling of the dilaton to the standard model.
Any theory leading to the same effective coupling will therefore
also lead to the same prediction for the masses.

In this note I will show that the dilaton can arise from the coupling of
the standard model to gravity, if a nonstandard regularization method
is adopted. Specifically, quantum fluctuations of the matter sector
induce a nontrivial effective potential for the conformal factor of
the metric, which then becomes an independent variable, the dilaton.
The effective potential of [4] can be derived in this way.
This idea has been studied before in the context of pure gravity [5];
very similar ideas have been discussed also in [6].
I will first illustrate the point in the case of a single real
scalar field, since this example already has all the features
of the more realistic models.

Let us therefore start from the action
$$
S(\varphi,g_{\mu\nu})=-\int d^4x\ \sqrt{g}
\left[{1\over2}g^{\mu\nu}\partial_\mu\varphi\partial_\nu\varphi
+{1\over2}(m^2+\xi R)\varphi^2+{\lambda\over 4!}\varphi^4\right]\ ,
\eqno(1)
$$
describing the coupling of the scalar field $\varphi$ to the metric
$g_{\mu\nu}$. We are interested in the effective dynamics induced
by the scalar field for the conformal factor, so we will restrict
our attention to metrics of the form
$$
g_{\mu\nu}=\rho^2{\bar g}_{\mu\nu}\ ,\eqno(2)
$$
where ${\bar g}$ is some fixed metric.
Defining $\phi=\rho\varphi$, the action (1) can be written as
$$
S(\phi,\rho,{\bar g}_{\mu\nu})=-\int d^4x\ \sqrt{\bar g}
\Bigl[{1\over2}{\bar g}^{\mu\nu}\partial_\mu\phi\partial_\nu\phi
+{1\over2}(m^2\rho^2+\xi\bar R)\phi^2+{\lambda\over 4!}\phi^4+\ldots\Bigr]\ ,
\eqno(3)
$$
where $\bar R$ is the Ricci scalar of the metric $\bar g$ and the
ellipses indicate terms containing derivatives of $\rho$.
This form of the action is invariant under the transformations
$$
{\bar g}'_{\mu\nu}=\omega^2{\bar g}_{\mu\nu}\ ,\qquad
\rho'=\omega^{-1}\rho\ ,\qquad
\phi'=\omega^{-1}\phi\ .\eqno(4)
$$
The classically equivalent action (1) is invariant under these transformations
in a trivial way, since the combinations $g_{\mu\nu}$ and $\varphi$
are not affected at all. Thus the symmetry (4) is a ``compensator'' or
St\"uckelberg type gauge invariance and there cannot be any anomaly for it,
as exhaustively argued in [7]. I will call the transformations (4)
``St\"uckelberg--Weyl'' transformations and reserve the name ``Weyl''
transformations for the conformal rescalings of the metric $g$.

In the case that the scalar field is free ($\lambda=0$, $\xi=0$),
the effective action $\Gamma(g_{\mu\nu})$ is minus one half of the
logarithm of the determinant of the operator
$\Delta_g+m^2$, where $\Delta_g$ is the covariant laplacian
$\Delta_g=-{1\over\sqrt{g}}\partial_\mu\sqrt{g}g^{\mu\nu}\partial_\nu$.
One expects from standard quantum gravity arguments that it
be of the form $\Gamma(g_{\mu\nu})=\int d^4x\ \sqrt{g}\left[\Lambda_{\eff}+
\kappa_\eff R+ O(R^2)\right]$.
In the case of a metric of the form (2) with ${\bar g}_{\mu\nu}$ flat,
the first term can be interpreted as an effective potential for the
conformal factor of the form $\Lambda_\eff \rho^4$. Invariance under
the transformations (4) implies that this is the only possible
form of the potential for $\rho$.

Let us now compute directly the one-loop effective action for
constant $\phi$ and $\rho$
by integrating over small fluctuations of the field $\phi$
in the action (3), with ${\bar g}_{\mu\nu}$ flat. It is given by
$$
\Gamma^{(1)}(\rho,\phi)=-{1\over2}\int d^4x \int {d^4q\over(2\pi)^4}
\ln(q^2+H) \eqno(5)
$$
where $q^2=\delta^{\mu\nu}q_\mu q_\nu$ and
$H=m^2\rho^2+{1\over2}\lambda\phi^2$.
One can now proceed as one would with any theory in flat space.
The integral can be regularized by imposing the cutoff $q^2<\Lambda^2$.
Adding suitable counterterms of the form $\Lambda^2\rho^2$
and $\ln\Lambda \rho^4$ one arrives at the renormalized one-loop
effective potential
$$
V^{(1)}(\rho,\phi)={1\over64\pi^2}
H^2\ln\left({H\over\mu^2}-{3\over2}\right)\ ,\eqno(6)
$$
where $\mu$ is a renormalization mass.

This effective potential is not simply of the form that one expected:
In particular when $\lambda=0$ it does not simply become quartic in $\rho$.
Also, invariance under the transformations (4) has been broken.
As explained in [5], this is due to the way in which the theory has
been regulated: the cutoff procedure used above introduces a
dependence on ${\bar g}$ which is not accompanied by a corresponding
factor of $\rho$. If we used the alternative cutoff
$g^{\mu\nu}q_\mu q_\nu=\rho^{-2}q^2<\Lambda^2$, the renormalized
effective potential would be purely quartic in $\rho$:
$$
V^{(1)}(\varphi,\rho)=
{1\over64\pi^2}h^2\ln\left({h\over\mu^2}-{3\over2}\right)\,\rho^4\ ,
\eqno(7)
$$
where $h=H/\rho^2=m^2+{1\over2}\lambda \varphi^2$.
Thus it is this second regularization method which yields
results in agreement with traditional quantum gravity.

The choice between the two regularization procedures is related to
which one of the metrics $g$ and ${\bar g}$ is interpreted as giving
the geometry of spacetime. In fact, the geometry enters in the
definition of the modulus squared of the momentum.
So if the geometry is given by $g$, one is
led to the effective potential (7), while if the geometry is given by
${\bar g}$ one arrives at the effective potential (6).

One may still worry that the logarithmic terms in (6) are an artifact
of the momentum cutoff regularization, and that they could not arise
if an invariant regularization was used. I will therefore now
rederive the effective potential (6) using the heat kernel regularization
in a curved background. This calculation is of independent interest
since it gives also the form of the curvature term in the effective action.

We begin from the linearization of the action (3)
$$
S_{\rm lin}(\phi,\rho,\bar g_{\mu\nu})=
{1\over2}\int d^4x\ \sqrt{{\bar g}}\,
\delta\phi\,\bo\,\delta\phi\ .
\eqno(8)
$$
where $\bo=\Delta_{\bar g}+\xi\bar R+H=\to+H$ and
$\Delta_{\bar g}$ is the covariant Laplacian in the metric ${\bar g}$.
The effective action is $\Gamma(\phi,\rho,\bar g)=-{1\over2}\ln\det\bo$.
The determinant can be defined through the formula
$$
\ln\det\bo=-\int_{1\over\Lambda^2}^\infty ds\, s^{-1}
{\rm Tr}\,e^{-s\bo} \ ,\eqno(9)
$$
where $\Lambda$ is an ultraviolet cutoff.
In order to extract the exact dependence of $\Gamma$
on constant $\rho$ and $\phi$ we write
$e^{-s\bo}=e^{-s\to}e^{-sH}$
and use the asymptotic expansion of the heat kernel
of $\Delta_\to$. Then the r.h.s. of (9) becomes
$$
-\int_{1\over\Lambda^2}^\infty\!ds\!
\int\, d^4x \sqrt{\bar g}\left(
b_0(\to)e^{-sH}s^{-3}
+b_2(\to)e^{-sH}s^{-2}
+b_4(\to)e^{-sH}s^{-1}
+\ldots\right)\ .\eqno(10)
$$
The integration over $s$ can be performed explicitly (see section 5 in [8]).
Using a suitable renormalization scheme
one arrives at the effective action
$$
\Gamma^{(1)}(\phi,\rho,\bar g)={1\over64\pi^2}\!
\int d^4x\ \sqrt{{\bar g}}\Bigl[
-H^2\!\left(\ln{H\over\mu^2}-{3\over2}\right)
+2\left({1\over6}-\xi\right)\bar R H\!\left(\ln{H\over\mu^2}-1\right)
+\ldots\Bigr]\, ,\eqno(11)
$$
where the ellipses stand for terms quadratic in $\bar R$.
We see that when ${\bar g}$ is flat the potential (6) is reproduced.

If the same calculation is performed starting from the
action (1), the relevant operator is $\o=\Delta_g+h$, and
one arrives at an effective
action $\Gamma(\varphi,g_{\mu\nu})$ which is identical to (11) except
for the replacement of $H$, $\bar g$ and $\bar R$ by $h$, $g$ and $R$.
In this effective action $\rho$ only appears within the
metric $g$. When $\bar g$ is flat and $\rho$ is constant, the
effective potential (7) is reproduced.

As in the case of the flat space calculations with cutoff, the
choice between the two quantization procedures depends on whether
$g$ or $\bar g$ is interpreted as the geometric metric.
In fact in the heat kernel regularization one isolates
the divergences by looking at the coincidence limit of a Green function
with the distance between the points
measured with respect to a certain geometry. In the first calculation,
this geometry was given by ${\bar g}$, and this led to the appearance
of a nontrivial potential for $\rho$, in the second calculation
it was given by $g$ and the effective potential is given just by
a cosmological term.

To summarize the discussion up to this point, we have seen that
when a scalar field is coupled to gravity, there exists a way of
quantizing the theory which produces a nontrivial effective potential
for the scalar and the conformal factor.
As a result, the invariance under the St\"uckelberg--Weyl transformations
(4) is broken. While the classical theory depends on the variables
$\phi$, $\bar g$ and $\rho$ only through the combinations $\varphi$
and $g$, in the quantum theory this is no longer the case.
The variable $\rho$, which classically can be eliminated,
becomes an independent dilaton field. This is similar to the effect
of the conformal anomaly [9], but it should not be confused with it.
(Note that the conformal anomaly is proportional
to curvature invariants, whereas the effect discussed here is
present already in flat space).
Strictly speaking, one should say that a theory is anomalous
only when there is no way to quantize it which preserves all classical
symmetries. We have seen that this theory is not anomalous.
Instead, we have used a quantization procedure that
``unnecessarily'' breaks the St\"uckelberg--Weyl invariance.
While it may seem more natural to choose the quantization procedure that
preserves the invariance, the alternative that is proposed here
does not seem at this stage to lead to any pathology and remains
a viable alternative.

The results obtained for the scalar field can be generalized to
other fields. A massive Dirac fermion coupled
to the scalar $\varphi$ has an action equal to
$$
\int d^4x\,\sqrt{g}\,\bar\chi (\dslash_g+m+h\varphi)\,\chi\ ,\eqno(12)
$$
where $\dslash_g=\gamma^a\theta_a{}^\mu D_\mu$ is the Dirac operator
in the metric $g$, $\theta_a{}^\mu$ is the inverse vierbein for
$g$ and $h$ is a Yukawa coupling. Defining
$\theta^a{}_\mu=\rho\,\bar\theta^a{}_\mu$ and $\psi=\rho^{3/2}\chi$,
the action can be rewritten as
$$
\int d^4x\,\sqrt{\bar g}\,\bar\psi\left(\dslash_{\bar g}
+m\rho+h\phi+\ldots\right)\psi\ ,\eqno(13)
$$
where the ellipses indicate terms involving derivatives of $\rho$.
The fermionic contribution to the effective action for
constant $\phi$ and $\rho$ is
$$
{1\over2}\ln\det(\Delta_{\bar g}+{1\over4}\bar R+F)
={4\over 64\pi^2}\int d^4x\,\sqrt{\bar g}
\left[F^2\left(\ln{F\over\mu^2}-{3\over2}\right)
+{1\over6}\bar R\,F\left(\ln{F\over\mu^2}-1\right)+\ldots\right],
\eqno(14)
$$
where $F=(m\rho+h\phi)^2$. Note that when $m=0$ the action is invariant
under rescalings of the metric and accordingly the contribution
to the effective potential is independent of $\rho$.

Similarly in the case of gauge fields, due to the Weyl invariance of
the Yang--Mills action, the effective potential depends only on scalar
fields and not on $\rho$. For example in the case of an abelian
gauge field coupled to a complex scalar $\phi$ the contribution to
the effective potential in the Landau gauge is
$$
-{1\over2}\ln\det(\to^\mu_\nu+\delta^\mu_\nu Z)=
{3\over64\pi^2}\int d^4x\,\Bigl[
-Z^2\left(\ln{Z\over\mu^2}-{3\over2}\right)
-{1\over6}\bar R\,Z\left(\ln{Z\over\mu^2}-1\right)+\ldots\Bigr]\, ,
\eqno(15)
$$
where $\to^\mu_\nu=\delta^\mu_\nu\Delta_{\bar g}+\bar R^\mu_\nu$
and $Z=g^2\phi^2$ is the effective mass squared of the gauge field.

In equations (11), (14) and (15) we have assumed that $\bar R$ is
small compared to $H$, $F$ and $Z$ respectively. Thus the terms
of order $\bar R$ represent the effect of a weak background
gravitational field on the effective potential.

Let us now come to the standard model. As in [1,2,4] we are going to
consider the possible presence of a term proportional to $\rho^4$.
We assume that the potential term in the action has the form
$-\int d^4x\sqrt{g}{\lambda\over4!}\left(|\varphi|^2-a^2\right)^2$,
where $\varphi$ is now a complex Higgs doublet.
This can be rewritten $-\int d^4x\sqrt{\bar g}V^{(0)}$, where
$$
V^{(0)}={\lambda\over4!}\left(|\phi|^2-\rho^2a^2\right)^2\ .\eqno(16)
$$
Putting together the contributions of the Higgs fields,
the gauge fields and the fermions, the one-loop effective
potential has the form
$$
\eqalign{
V^{(1)}={1\over 64\pi^2}\Biggl[&
H^2 \left(\ln {H\over \mu^2}-{3\over 2}\right)
+3G^2 \left(\ln {G\over \mu^2} -{3\over 2}\right)\cr
&+6W^2 \left(\ln {W\over \mu^2}-{3\over 2}\right)
+3Z^2 \left(\ln {Z\over \mu^2}-{3\over 2}\right)
-12T^2 \left(\ln {T\over \mu^2}-{3\over 2}\right)\Biggr]\, , \cr}\eqno(17)
$$
where
$$
\eqalign{
H&={\lambda\over6}(3\phi^2-a^2\rho^2)\ ,
\qquad G={\lambda\over6}(\phi^2-a^2\rho^2)\ ,\cr
W&={1\over 4}g_2^2 \phi^2 ,\qquad Z={1\over 4}(g_2^2 +g_1^2)\phi^2 ,
\qquad T={1\over 2}h^2 \phi^2. \cr}\eqno(18)
$$
One can use the potential $V^{(0)}+V^{(1)}$ and the known values of
$\phi^2$, $W$ and $Z$ to rederive the prediction of the top and Higgs
masses that was given in [4].
The one-loop term (17) differs from the result given in [4] in the
numerical coefficients of the gauge field contributions.
This is due to the different renormalization conditions and should be
irrelevant if the appropriate values of $\phi^2$, $W$ and $Z$ are used.

We now discuss the induced action for the metric $\bar g$.
The coefficient of $\sqrt{\bar g}\bar R$ in the effective action
is equal to
$$
\eqalign{
{1\over 64\pi^2} \Biggl[&
{1\over3}H\left(\ln{H\over \mu^2}-1\right)
+G\left(\ln{G\over \mu^2}-1\right)\cr
&-W\left(\ln{W\over \mu^2}-1\right)
-{1\over2}Z\left(\ln{Z\over \mu^2}-1\right)
+{2\over3}T\left(\ln{T\over \mu^2}-1\right)\Biggr]\ .
\cr}\eqno(19)
$$
Thus the induced action for $\bar g$ is not simply an
Einstein--Hilbert term. Rather, we have a complicated
interaction between the Higgs field, the dilaton and the metric,
which is more similar to a Brans--Dicke type of action.
However, at momentum scales much smaller
than the Higgs mass, the coefficient of
$\bar R$ can be treated as essentially constant,
thus leading to an Einstein-Hilbert effective action for $\bar g$ [8].
Unfortunately, this cannot be the whole story, since the
v.e.v. of (19) is of the order of the $W$ mass squared and thus cannot be
identified as the coefficient of $\bar R$ in the whole gravitational
action.

If the Einstein-Hilbert term is to be obtained as an induced action,
it must receive contributions from other sectors of the theory.
I will briefly discuss here the possible contribution of the gravitational
sector, which was computed in [5] starting from a gauge theory of
gravity with lagrangian quadratic in curvature and torsion.
There is no potential at tree level, and the one-loop term
depends only on $\rho$:
$$
V^{(1)}={9\over64\pi^2}e^4\rho^4\left(\ln{e^2\rho^2\over\mu^{'2}}
-{1\over2}\right)\ ,\eqno(20)
$$
where $e$ is the gauge coupling constant of the Lorentz connection
and $\mu'$ is another renormalization mass.
Similarly, the coefficient of $\bar R$ will be of the order of
$e^2\rho^2$. We can therefore interpret the induced $\bar R$
term as the whole gravitational action if the v.e.v. of $\rho$
is of the order of Planck's mass.

In principle the v.e.v of $\rho$ is the minimum of the total potential
obtained summing (16), (17) and (20). However, the natural way to obtain
v.e.v.s for $\rho$ and $\phi$ of the order of the Planck
and Fermi scale respectively, is to have the coupling constant $a$
very small. Then, the dependence on $\rho$ in (16) and (17) can be
neglected with respect to (20) and the expectation value of $\rho$ will be
essentially equal to $\mu'/e$, the minimum of (20).
One can then insert this value in (16) and (17) and minimize with
respect to $\phi$; the result will be of the order of $a\rho$ and
this will fix $a$ to be of the order of $10^{-16}$.
Therefore in this picture, the large hierarchy between the Planck
and the Fermi scale would be due to the smallness of the coupling
between Higgs and dilaton.
Clearly a renormalization group analysis will be needed
to make further progress in this direction.
In contrast to [1,2,4], the dilaton would have
a mass of the order of Planck's mass.

\bigskip
\centerline{\bf Acknowledgements}
\smallskip
\noindent
It is a pleasure to thank R. Floreanini and L. Griguolo for
important contributions to the point of view presented here.
I also wish to thank G. Horowitz, J. Madore and R. Woodard for
conversations and comments. This work is supported by INFN
and also by the National Science Foundation under grant
No. PHY89-04035.
\bigskip
\centerline{\bf References}
\medskip
\noindent
\item{[1]} R.D. Peccei, J. Sol\`a and C. Wetterich,
Phys. Lett. B {\bf 195}, 183, (1987);\hfil\break
C. Wetterich, Nucl. Phys. {\bf B 302}, 668 (1988);\hfil\break
S.M. Barr and D. Hochberg, Phys. Lett. B {\bf 211}, 49 (1988);\hfil\break
G.D. Coughlan, I. Kani, G.G.Ross and G. Segr\`e,
Nucl. Phys. {\bf B 316}, 469 (1989);\hfil\break
E.T. Tomboulis, Nucl. Phys. {\bf B 329}, 410 (1990).
\smallskip
\item{[2]} W. Buchm\"uller and N. Dragon, Phys. Lett B {\bf 195}, 417 (1987);
\hfil\break
W. Buchm\"uller and N. Dragon, Nucl. Phys. {\bf B 321}, 207 (1989);
\hfil\break
W. Buchm\"uller and C. Busch, Nucl. Phys. {\bf B 349}, 71 (1991).
\smallskip
\item{[3]} A. Chamseddine, G. Felder and J. Fr\"ohlich, ``Gravity in
non-commutative geometry'', ZU-TH-30/1992
(to appear in Comm. Math. Phys).
\smallskip
\item{[4]} A. Chamseddine and J. Fr\"ohlich, ``Constraints on the Higgs and
top quark masses from effective potential and noncommutative geometry'',
ZU-TH-16/1993.
\smallskip
\item{[5]} R. Floreanini and R. Percacci, ``Average effective potential for
the conformal factor'', SISSA 71/93/EP, hep-th/9305172.
\smallskip
\item{[6]} E. Elizalde and S.D. Odintsov, ``Gravitational phase
transitions in infrared quantum gravity'', Hiroshima preprint HUPD-92-10;
\hfil\break
A.A. Bytsenko, E. Elizalde and S.D. Odintsov, ``The renormalization
group and effective potential in curved spacetime with torsion'',
HUPD-93.
\smallskip
\item{[7]} N.C. Tsamis and R.P. Woodard, Ann. of Phys. {\bf 168}, 457
(1986).
\smallskip
\item{[8]} R. Floreanini and R. Percacci, Phys. Rev. D {\bf 46}, 1566 (1992).
\smallskip
\item{[9]} I. Antoniadis and E. Mottola, Phys. Rev. D {\bf 45}, 2013
(1992).

\vfil
\eject
\bye